\documentclass[aps,twocolumn]{revtex4}
\usepackage{times}
\begin{document}

\title{\LARGE\rm Information Processing beyond Quantum Computation%
\footnote{Keynote talk presented at the First World Congress on Lateral
Computing (WCLC 2004), Bangalore, December 2004, {\tt quant-ph/0306158}.}
}
\author{Apoorva Patel}
\affiliation{\rm Centre for High Energy Physics and
Supercomputer Education and Research Centre\\
Indian Institute of Science, Bangalore-560012, India\\
E-mail: adpatel@cts.iisc.ernet.in}
\maketitle

{\bf\small Abstract--
Recent developments in quantum computation have made it clear that there
is a lot more to computation than the conventional Boolean algebra.
Is quantum computation the most general framework for processing information?
Having gathered the courage to go beyond the traditional definitions,
we are now in a position to answer: \emph{Certainly not}.
The meaning of a message being ``a collection of building blocks''
can be explored in a variety of situations.
A generalised computational framework is proposed based on group theory,
and it is illustrated with well-known physical examples.
A systematic information theoretical approach is yet to be developed
in many of these situations.
Some directions for future development are pointed out.}

\section{\rm Motivation}

The silicon transistor was invented about half a century ago.
Since then the semiconductor technology has grown at a rapid pace to
pervade almost all aspects of our lives.
This growth has been so explosive---doubling the number of transistors
on a chip every 18-24 months according to Moore's law---that many choices
made in constructing the theoretical framework of computer science
(see for example, \cite{neumann}) were almost forgotten.
Computer architecture became essentially synonymous with digital electronic
circuits implementing Boolean operations, pushing aside other competing models.
Developments in quantum computation during the past decade have led us to
question this attitude, and brought in focus the fact that there is much
more to information theory than just Boolean logic.
The concept of what is computable and what is not has not changed,
but the criteria determining how efficiently a computational task
can be implemented have been altered.
The reason behind this change is that some of the implicit assumptions
of theoretical computer science are too restrictive,
when compared to physically realisable models.
Computational power of a framework can be enhanced by discarding such
unnecessary assumptions.
My aim in this article is to describe possible directions
of expansion for generalised information theory,
based on many physical examples we encounter in the world around us.

To obtain a perspective of how profoundly a subject can expand
up on modification of its fundamental postulates,
it is worthwhile to look at what happened in geometry.
Euclid's formulation of planar geometry had five postulates.
The fifth of these postulates concerned the number of lines
that can be drawn parallel to a given line
and passing through a point outside the given line.
Euclid considered one parallel line to be the self-evident answer,
but could not prove that.
So he included this property as a postulate in his theory,
even though it was quite different from the first four
which defined basic components of geometry.
This state of affairs no doubt troubled mathematicians.
For many centuries, they tried to deduce the fifth postulate
from the other four, and failed.
Only after two thousand years, they found the courage to discard this
fifth postulate, which resulted in development of curved space geometry.
We now know that a surface of positive curvature allows no parallel lines,
while a surface of negative curvature allows an infinite number of parallel
lines (if two parallel lines can be drawn passing through a point,
there will be an infinite number of parallel lines in between).

When the concepts of a subject are generalised,
it is quite common that definitions are extended,
new terminology is invented, and meanings of some old words change.
(For example, a line in a curved space has to be interpreted as a geodesic,
i.e. the shortest curve connecting two points.)
Such a situation can create unnecessary confusion,
while developing new methodology.
To guard against it, words used in technical contexts should be interpreted
only in technical terms, and not confused with their common language meanings.
I begin by describing my interpretation of some of the technical words
that appear in information theory.

\section{\rm Technical Concepts}

\begin{figure}
\begin{picture}(200,100)(0,0)
\put(60,28){\framebox(60,40){\shortstack{Physical\\ Device}}}
\put(0,45){Input}
\put(30,48){\vector(1,0){30}}
\put(160,45){Output}
\put(120,48){\vector(1,0){30}}
\put(70,90){Instructions}
\put(90,88){\vector(0,-1){20}}
\put(50,0){Oracles/Look-up Tables}
\put(90,24){\vector(0,1){4}}
\multiput(90,8)(0,4){4}{\line(0,1){2}}
\end{picture}
\vspace{2mm}

Fig. 1. Schematic representation of a computer.\\
Every computation may not use oracles or look-up tables.
\end{figure}

Information theory deals with two broad areas, communication and computation.
Communication is quite simple---the receiver gets
whatever is sent by the sender.
There is no processing in the middle, unless there is some unexpected noise.
Computation is more complex---the input is intentionally manipulated
by an external agency to produce an output.
(A schematic representation of a computer is shown in Fig. 1.)
The words data, information and knowledge often appear
in discussions of communication and computation.
I assign them specific meanings as explained below.

Data list physical properties of a system.
They describe a particular realisation of the physical system,
amongst its many possible states.
Data are often obtained by experimental observations of the system,
and generally provide the starting point of a computational process.
I stress that data are always firmly rooted in physical characteristics,
and should not be separated from them.

Information is the abstract mathematical property
obtained by detaching all the physical characteristics from data.
It just becomes a measure of the number of possible states
of the system \cite{shannon}.
This mathematical abstraction proves to be very useful,
because in dealing with information, at no stage one has to worry about
where the information came from or what it means.
The physical realisation of information may change according to
the convenience of the task to be carried out.
(For example, our electronic computers compute using electrical signals,
but store the results on the disk using magnetic signals;
the former realisation is suitable for fast processing,
while the latter is suitable for long term storage.)
Abstract information theory also allows manipulation of information
without going into nitty-gritty of its meaning,
e.g. compress data, quantify error rate, devise codes, and so on.

While extraction of abstract information from data allows one to
formulate precise mathematical rules for its systematic analysis,
the abstraction also brings in a limitation.
Though the manipulations of information can be defined as mathematical
algorithms, they have to be implemented by physical devices.
In order to manipulate information,
one must map information to physical properties,
and the types of manipulations that can be carried out
are limited by the types of physical devices available.
(For example, we use various programming languages to implement
mathematical algorithms on a computer.
On the other hand, the electronic computer hardware responds
only to voltages and currents.
So a whole hierarchy of translation machinery is constructed,
involving compilers and operating systems,
to convert the algorithms to binary machine codes
and then map them on to off/on states of silicon transistors.)
An important consequence of this physical dependence is that
the efficiency of a computational task cannot be determined solely by
its mathematical algorithm---the efficiency depends on the algorithm as
well as on the properties of the physical device that implements it.

The role of physical properties is also inevitable in adding
a sense of purpose to information, and converting it to knowledge.
If the receiver does not understand the language of the message,
he will just have random looking symbols and no meaningful interpretation.
He will gain the knowledge contained in the message,
only when he figures out the language.
As a matter of fact, the whole subject of cryptography
is based on sending the information but concealing its language.
A common language can be established between the
sender and the receiver only by physical means.
Of course, once a common language is established,
it can be used repeatedly in an abstract manner.
(For example, to teach a baby what a book is, we first show him the
book---perhaps tap it a few times while saying the word ``book''.
Once the association between the word and the object is established in the
baby's mind, we can reduce the physical dependence---point at the book from
afar, show a picture of it, and ultimately just utter or write the word.)
In case of the most primitive (or low level) messages,
there is no luxury of abstract languages---the only language that exists
is the one labeled by physical properties.
In such cases, the physical objects that carry the message have to convey
the information as well as its interpretation to the receiver.
Once again, an optimal language can only be designed
if the available physical means are known.

To summarise, \emph{data is not information and information is not knowledge}.
We have instead,\\
\phantom{abcdefg} Information = Data - Physical Realisation ,\\
\phantom{abcdefg} Knowledge = Information + Interpretation .\\
Abstract information theory does not tell us
what physical realisation would be appropriate for a particular message,
nor does it tell us the best way of implementing a computational task.
To make such choices, we must look at the physical resources available,
i.e. analyse the type of information and not the amount of information.

The number of fundamental physical interactions is rather small,
and that limits the possible physical realisations of a computer.
A variety of computational schemes can still be created, however,
by combining the fundamental ingredients in different ways.
To get an idea about the multitude of physical resources that can be used
to process information, it is instructive to look at biological systems.
Over billions of years, evolution has had plenty of time (which we do not
and cannot have) for experimentation with a wide range of physical systems.
Let us look at an example.

\section{\rm An Example from Biology}

Consider the following biological communication system
devoid of human involvement,
to get a feeling for the wide applicability of information theory.
A plant attracts an insect to its flower.
How does this take place?
The flower releases certain fragrant molecules
which are received by the insect.
How does the insect decide which molecule is fragrant and which is not?
That information is encoded in the 3-dimensional
atomic structure of the molecule,
which determines how it will bind to the smell-receptors of the insect.
How does the plant know which molecule to release,
and the insect know which molecule to look for?
That information resides in their genomes, which have evolved together
for millions of years and converged to a common language.
This convergence has a mutually beneficial purpose;
the plant gets pollinated and the insect obtains nectar as food.

It is even more fascinating to observe
how the insect finds the direction to reach the flower.
Neither does the flower know the location of the insect,
nor does it encode its own location
in the individual molecules of fragrance that it releases.
(Note that a predator uses the same communication scheme to hunt its prey.)
Instead, the flower releases not a single molecule of fragrance
but millions of them.
These molecules are identical and spread out in all possible directions.
The insect moves towards the flower by detecting
in which direction the concentration of molecules increases.
It finds the concentration gradient by random movement
as well as by using multiple receptors (which can detect parallax).
Since millions of messages are broadcast in this communication system,
the messages have to be produced cheaply to be energy efficient.
By going down all the way to the molecular scale,
the plant has indeed optimised and made every message quite cheap.

The features of this communication system,
especially the structure of the message and massive redundancy,
are in total contrast to the conventional implementation of
information theory and its optimisation criteria in computer science.
Yet there is no doubt that there is a purposeful communication
between the plant and the insect.
If we want an information theory capable of dealing with such
unusual types of communications, we must generalise its postulates
and expand its definitions beyond what is there in the textbooks.

\section{\rm Physical Criteria}

To expand the scope of information theory, I generalise the notion of a
message from ``a sequence of letters'' to ``a collection of building blocks''.
Collections can be labeled according to the number of external
space-time dimensions where the building blocks are arranged.
The building blocks themselves can be characterised by their properties,
which may be external or/and internal.
As already emphasised, the appropriate building blocks and collections
for a given information processing task have to be selected
based on physical principles.
Furthermore, the selection can be optimised depending on
what is available and what is to be accomplished.
The most efficient computers are those that
reliably accomplish their tasks using the least amount of resources.
The optimisation process is thus guided by two principles:
(i) minimisation of errors, and (ii) minimisation of physical resources.
These principles often impose conflicting demands,
and one has to learn how to tackle them in the process of computer design.

Laws of thermodynamics imply that unwanted disturbances can never be
completely eliminated---errors are an unavoidable fact of life.
So we must develop strategies to keep the error rate in control.
The system can be protected from external disturbances by shielding.
On the other hand, the system can be guarded against internal fluctuations
only if the information processing language is based on discrete variables
(as opposed to continuous variables).
Allowed values of fundamental physical variables are often continuous,
in which case a set of non-overlapping neighbourhoods
of discrete values can be chosen as the discrete variables.
The advantage is that the discrete variables remain unaffected,
even when the underlying continuous variables drift,
as long as the drifts keep the values within the assigned neighbourhoods.
This is the common procedure of \emph{digitisation}, it eliminates small
fluctuations and leads to the framework of \emph{bounded error computation}.
(For example, my handwriting is not the same as yours,
nor is my accent the same as yours.
Yet you can figure out what I write or what I speak,
because the letters and sounds of our languages are discrete.
A close match---and not an exact match---is sufficient for you
to understand what I convey.)

In a language based on continuous variables, it is not possible to tell
apart what is unwanted noise and what is a genuine transformation.
On the other hand, in a digitised language, all small fluctuations are
interpreted as unwanted noise, and are eliminated by resetting the
variables to their discrete values once in a while.
All large changes are interpreted as genuine transformations,
and so large erroneous changes still persist in a digitised language.
Digitisation is thus worthwhile, when large erroneous changes are rare.
In fact, large erroneous changes can be eliminated too,
provided their rate falls below a certain threshold,
with the help of error correcting codes based on redundancy and nesting.
(For example, we rather unconsciously change our adult language
when talking to babies.
The baby language has less number of sounds and its words are full of
repetitive sounds---a simple error correction procedure in a situation
of high transmission loss.)

It is useful to note that quantum physics at the atomic scale
automatically provides discrete variables,
e.g. finite size of atoms leads to lattices,
and discrete energy levels lead to characteristic transitions.
In other cases, there is a loss of precision when changing from
continuous variables to discrete ones, e.g. discrete variables can
produce integers and rational numbers but not irrational numbers.
Yet the framework of bounded error calculations is immensely useful,
because in all our practical applications we never need results with
infinite precision; as long as results can be obtained within a
a prespecified non-zero tolerance limit, they are acceptable.
(For example, a wheel does not have to be exactly circular to be useful;
all we require is that it should be round enough to roll.)
The error rate depends on the physical device processing information,
and the tolerance limit is specified by the computational task to be
carried out---bringing them together is a question of computer design.

Physical resources to be optimised include space, time and energy.
Minimisation of spatial resources means carrying out
the computational task using as few physical components as possible,
e.g. memory and disk space in our digital computers.
In addition to finding a software algorithm which requires the smallest
number of variables, this also requires selecting elementary hardware
components that are simple and easily available,
and yet versatile enough to be connected together in many different ways.
This is the common choice at the lowest level of information processing,
and complicated systems are then constructed by packing a large number
of components in a small volume.
Correlations and repetitive structures in a language waste spatial
resources---periodic crystals are no good; information content
of a language resides in its aperiodic random patterns.
The language is most versatile when its building blocks
can be arranged in as many different ways as possible.
(For example, this is what one exploits when compressing files in a computer.)

Minimisation of temporal resources means finding an algorithm with the
smallest number of execution steps, and also finding hardware components
that allow fast implementation of computational instructions.
Often a trade-off is possible between spatial and temporal resources,
and specific choices are made depending on what is more important,
e.g. parallel computers save on time by using more hardware.

A computer is a driven physical system,
with irreversible operations of resetting and erasure.
So, according to thermodynamical laws,
a source of free energy is required to run it.
This thermodynamical limit is not of much practical relevance, however,
because available physical devices are nowhere near that efficiency.
Energy consumption during information processing depends almost
entirely on the choice of hardware technology.
The best strategy is to make the hardware components as tiny and as cheap
as possible, so that they can carry out their tasks consuming little energy,
and also recycle energy wherever possible.

Now we can see that conflicts arise amongst these optimisation guidelines.
Tiny components and fast operations are less reliable and increase noise,
error correction procedures add overheads to physical resources,
more precise operations demand more energy, segregating different
ingredients of a computational task and assigning them to specialised
components increases the reliability of computation but increases
resource requirements, and so on.
There is no easy way to figure out the optimal language for a given 
computational task.
Depending on how much weight is assigned to which criterion,
different languages can be designed to implement the same computational task.
We know by experience that when the languages are versatile enough,
information can be translated from one language into another by replacing
one set of building blocks and operations by another set of building blocks
and operations.
Subjective (and historical) choices have often dictated specific realisations.

When a number of choices are available, the language with the smallest set
of building blocks has a unique status in the optimisation procedure:\\
(a) Generically, physical hardware properties have a fixed range of values.
Decreasing the number of discrete values allows them to be put
as far apart from each other as possible within that range.
This dispersal minimises misidentification,
and provides the largest tolerance against errors.
(For example, silicon transistors are powerful non-linear electrical devices,
but they are used in digital computers only as two extreme saturated states.)\\
(b) Reduction of possible physical states of elementary components
simplifies the instruction set needed to manipulate them,
and also the possible types of connections amongst the components.
(For example, with our decimal number system, we had to learn $10\times10$
tables in primary schools to do arithmetic.
With the binary number system,
our computers implement the same arithmetic with only two Boolean operations,
XOR for addition and AND for multiplication.)\\
(c) A small number of discrete states increases the depth of computation,
i.e. the number of building blocks required to represent a fixed amount
of information.
But with only a small number of states and instructions,
elementary components can be made small and individual instructions fast.
Typically, high density of packing and quick operations
more than make up for the increase in the depth of computation,
and the overall requirement for physical resources goes down.\\
(d) At the lowest level of information processing,
translation of languages is not possible, and only a handful of
instructions related to physical responses of the hardware exist.
The simplest language is then a distinct advantage,
and it becomes the universal language for that particular hardware.

\section{\rm Types of Collections}

We are now in a position to look at some examples of information processing
systems, and understand how well they implement the optimisation principles. 
Messages are constructed by linking the basic components---the building
blocks of the language---in a variety of arrangements.
The information contained in a message depends
on the values and positions of the building blocks.
Any language that communicates non-trivial information
must have the flexibility to arrange its building blocks
in different ways to represent different messages.
Any physical realisation of the message must involve
physical phenomena to put the building blocks together.

Let us look at possible collections of building blocks:\\
$\bullet$ \emph{0-dim:}
Such a collection requires multiple building blocks
to be at the same point in space and time.
This is the phenomenon of superposition,
which is a generic property of waves.
Superposition allows many signals to be combined together,
and then also be manipulated together, but at the end
only one of the signals can be extracted from the collection.
(For example, radio and television broadcasts combine multiple
electromagnetic signals together, and the receiver extracts the desired
signal---only one at a time---by tuning to the corresponding frequency.)\\
$\bullet$ \emph{1-dim:}
Here the message is an ordered sequence of building blocks.
This is the most common form used in conventional information theory.
Mathematically, the collection is expressed as a tensor product
of individual components.
The ordering of the sequence can be either in space or time,
e.g. our written and spoken languages.\\
$\bullet$ \emph{2-dim:}
Higher than one dimensional collections can be viewed
as combinations of multiple ordered sequences.
The simplest situation is that of parallel computation,
based on multiple similar information processing units.
Such parallelism allows an unusual feature,
namely information can reside in correlations amongst sequences
without being present in any individual sequence.
Biological systems have efficiently exploited this feature,
whereby gradients are detected at the cost of redundancy.
(For example, multiple detectors are commonly used to estimate distance,
either by parallax removal or by detecting concentration
gradients---the former uses waves while the latter uses particles.)
Such systems have been left out of our computers,
and our computers are not at all efficient at finding gradients.
We are gradually learning to use such systems for certain tasks,
e.g. very long base-line interferometry (VLBI) in astronomy,
global positioning system (GPS) in geography,
and space-time codes in electronic communications.\\
$\bullet$ \emph{3-dim}:
Such collections describe the physical structure of an object
in our three dimensional space.
Structural information is useful for establishing lock-and-key mechanisms
that can trigger an appropriate response.
(For example, proteins use such a system to carry out various tasks in
living organisms.)\\
$\bullet$ \emph{4-dim}:
This would be a complete description of any event, either past or future,
in our universe with one time and three space dimensions.
Such a description would contain all the information about a system,
that can ever be extracted.
On the other hand, it is too much for our common use,
and we typically use only a smaller dimensional subsystem for our tasks.

It is not necessary that a collection of building blocks
be restricted to a fixed dimensionality.
In fact, computational capability of a system can be vastly enhanced
by simultaneous use of features of different dimensionalities.
For example, the framework of quantum computation \cite{nielsen}
uses collections of both zero and one dimension.
The phenomenon of superposition, combined with the ordered sequence of qubits,
leads to the unusual possibility of quantum entanglement of states.
It is this combination which enables a quantum computer to solve certain
problems much more efficiently compared to a classical computer.

Another example of multiple dimensionality is provided by proteins,
which possess features of both one and three dimensional collections.
The one dimensional form of proteins is convenient
for efficient synthesis through polymerisation of amino acids,
and also for crossing cellular membranes through narrow channels.
The three dimensional form is suitable for carrying out various functions
through highly selective binding to other molecules of complementary shape.
The mechanism for realising both these forms is based on the property
that proteins are physical systems poised at the edge of criticality.
Small changes in suitable external parameters
(e.g. concentration of a denaturant or pH of the solvent)
can unfold the protein to its polypeptide chain form,
or conversely, fold it into its three dimensional native form.

Superposition, parallax, phase transitions,
are all well understood physical phenomena.
The examples above illustrate how the capability of an information
processing system can be enhanced by incorporating them in physical devices.
Our conventional framework of computation has barely made a start in that
direction.

\section{\rm Types of Building Blocks}

Physical properties of building blocks, in both internal and external space,
are generically organised in terms of groups.
(There is an implicit assumption here that
we can recognise the same object in different manifestations,
just as we can identify the same person wearing different clothes.)
As discussed above, for a given information processing task,
the smallest discrete group that can implement it
is the ideal candidate for the optimal language.
When the group of physical properties is a continuous one,
we must look for its smallest yet faithful discrete subgroup.
We have also observed earlier that because of unavoidable noise,
a discrete building block of a language is associated not with just
a point on the group manifold but with a neighbourhood of the point.
Thus to specify the building blocks completely,
we have to describe neighbourhoods of discrete group elements.

The algebra of any group is fully specified in terms of its generators.
The number of independent generators gives the dimensionality of the group.
In case of continuous groups, the generators define a vector space.
In a $d$-dimensional group manifold,
any group element is specified by $d$ coordinates.
One more parameter is needed to specify the neighbourhood of a group element.
For generic manifolds, such a $(d+1)$ parameter object is called a simplex.
It provides the simplest specification of an elemental group volume
which faithfully realises all group properties.
The smallest discrete realisation of any group, therefore,
corresponds to replacing the entire group by a single simplex.

Sometimes the dual (Fourier) space of representations provides a more convenient
description of the group than the coordinates specifying the group elements.
In that case, the minimal set of $(d+1)$ elementary building blocks
is formed by the $d$-dimensional fundamental representation
and the $1$-dimensional identity representation.
Any other representation of the group can be obtained by putting together
several of these elementary building blocks.

In general, the building blocks are completely characterised in terms of
two discrete groups, one for the external properties and one for the
internal ones (one of the groups may be trivial in some cases).
Let us look at the minimal set of building blocks for some common groups:\\
$\bullet$ \emph{1-dim:}
Groups with a single generator include cyclic groups,
the set of integers and the real line.
The minimal simplex in this case has just two points, $Z_2=\{0,1\}$.
It forms the basis of Boolean arithmetic widely used in digital computers.
The binary language can be easily extended to a $d$-dimensional situation,
as the Cartesian product $(Z_2)^d$, and is therefore convenient
as a general purpose language in handling a variety of problems.\\
$\bullet$ \emph{2-dim:}
The simplex for two dimensional geometry is a triangle.
Triangulation is useful in discrete description of arbitrary surfaces.
At nano-scale, its dual hexagonal form  can be realised
in terms of the $sp^2$-hybridised orbital structure of graphite sheets,
which may become useful in lithographic techniques.\\
$\bullet$ \emph{3-dim:}
In three dimensional space, the simplex is a tetrahedron.
At molecular level, $sp^3$-hybridised orbitals provide its dual form.
Arbitrary structures can be created by gluing tetrahedra together.
Tetrahedral geometry based on properties of carbon provides a convenient
starting point for understanding the three dimensional language of
proteins \cite{patel}.\\
$\bullet$ $SU(2):$
This is also a group with three generators,
up on which description of quantum bits is based.
Arbitrary states of a qubit, including the mixed states arising from
decoherence (i.e. environmental noise), can be fully described using
a density matrix, which is a linear combination of four operators
$\{1,\sigma_x,\sigma_y,\sigma_z\}$.

Larger groups have been used in error correcting codes and cryptography,
but not for processing information.

\section{\rm Types of Processing}

Once the physical properties of the building blocks are fixed,
i.e. the discrete groups describing their external and internal properties, 
the possible computational operations are just group transformations.
Different physical means are needed in case of different groups,
and what is possible and what is not depends on the available technology.
Nonetheless, it is straightforward to list the possibilities:\\
$\bullet$ \emph{0-dim:}
The only mathematical operation allowed with superposition is addition.
Addition is commutative,
and interference effects produced by it are computationally useful.\\
$\bullet$ \emph{1-dim:}
This is the most common realisation, where two different
group operations of addition and multiplication are possible.
Both operations are commutative, their combination obeys
a distributive rule, and all our arithmetic is based on them.
In mathematical terms, $Z_2$ is a field---the smallest one.\\
$\bullet$ \emph{dim$>$1:}
In higher dimensions, addition generalises to translation.
The obvious generalisation of multiplication is scale transformation,
but the scope of multiplication can be expanded to include rotations
as well (which can be viewed as multiplication by a matrix).
Rotations are commutative in two dimensions, but non-commutative for $d>2$.
Discrete operations of translation, rotation and scale transformation
can be realised on a lattice made of simplicial building blocks.
The algebra generated by them is much more powerful than common arithmetic.

Clearly, more and more group operations become possible
as the dimensionality of the group increases.
Direct physical implementation of a complicated group operation
can substantially reduce the depth of computation.
For example, steps of a quantum algorithm can be represented
in classical language as multiplication of unitary matrices
with superposed state vectors.
Such a multiplication is a single operation on a quantum computer,
but an elaborate procedure on a classical computer,
and therein lies the physical advantage of a quantum computer.
From this point of view, we have hardly begun to explore
the power of non-commutative group algebra.

\section{\rm Future Outlook}

We are accustomed to looking at our computers from the top level
down---from the abstract mathematical operations to the transistors
embedded in silicon chips.
On the other hand, to be able to design efficient computers,
we must study them from the bottom level up---from
the elementary building blocks to the complicated languages.
Biological systems are a useful guide in such an exercise,
because they have indeed evolved in that manner,
from biomolecular interactions to multicellular organisms,
and have explored a variety of options along the way.

We have seen that the scope of ``information processing'' can be vastly
enhanced by looking at a message as a ``collection of building blocks''.
Physical optimisation criteria require discrete languages, versatile
operations, special purpose components and tiny building blocks.
But beyond that, there is lot of freedom in the choice of building blocks.
A variety of computational frameworks can be constructed by appropriate
choice of (i) the dimensionality for the arrangement of building blocks,
and (ii) the group structure for the properties of building blocks.
I have described several physical computational systems above,
and pointed out the choices inherent in their design.
It is natural to look for other possible choices, which may help
in finding the optimal hardware design for a given computational task,
and which may lead to novel computational schemes:\\
$\bullet$ Operations of calculus, such as differentiation and integration,
are easier to carry out using continuous variables instead of discrete ones.
Although digitisation is necessary to control errors,
it does not have to be imposed at every computational step.
So the framework of analogue computation, punctuated by digitisation,
may turn out to be convenient for implementing operations of calculus.\\
$\bullet$ The depth of computation can be reduced by direct execution of
complex high level instructions (i.e. without translation to lower levels).
This can be achieved using special purpose components and configurable
systems.
In fact, such features are commonplace in biological systems.\\
$\bullet$ A fractal arrangement would be an unusual collection of building
blocks.
Such self-similar patterns occur in concatenated error correcting codes,
but can they be useful in some new type of information processing?\\
$\bullet$ Use of building blocks having multiple physical properties,
each described by a particular group, can cut down resource requirements
by simultaneous execution of multiple transformations.
Such physical objects exist,
e.g. an electron has location, spin, energy level etc.,
and quantum computation has provided the first step in this direction.\\
$\bullet$ Use of large groups can also reduce depth of computation.
Such groups have been used in cryptography, but can we design physical
building blocks that directly implement them?\\
$\bullet$ $(Z_2)^d$ does not provide the minimal set of building blocks for
$d>1$; it contains $2^d$ points compared to $(d+1)$ points of a simplex.
A simplicial geometry can be more efficient for multi-dimensional
information processing.

Construction of the complete information theory framework
for a general set of building blocks is a wide open subject.
The mathematical definition of information parallels
the thermodynamical definition of entropy.
Entropy just counts the number of available states,
and there is no hurdle in applying it
to configurations of arbitrary building blocks.
The quantification of correlations amongst the building blocks, however,
becomes increasingly complicated as the dimensionality increases.
We have made a start in this direction, in our efforts to incorporate
superposition and entanglement in quantum information theory,
where Boltzmann entropy is generalised to von Neumann entropy.

Indeed, there is a lot to explore in both computational hardware and
software---methods to deduce optimal physical realisation of building blocks,
and group theoretical techniques to construct high level instructions,
would be inseparable in that.

\end{document}